\documentclass[11pt]{article}
\usepackage{units}
\usepackage[colorlinks]{hyperref}
\usepackage{threeparttable}
\usepackage{subfigure}
\usepackage{booktabs}
\usepackage{multirow}
\usepackage{amsmath}
\usepackage{amsfonts}
\usepackage[left=1.3cm, bottom=1.3cm, top=1.2cm, right=1.2cm]{geometry}
\usepackage{tikz}
\usetikzlibrary{shapes.geometric}

\date{November 23, 2022}

\usepackage{enumitem}
\setitemize{topsep=3pt, itemsep=1pt, parsep=2pt, leftmargin=1.5em}
\setenumerate{topsep=3pt, itemsep=1pt, parsep=2pt, leftmargin=1.5em}
\setdescription{topsep=3pt, itemsep=1pt, parsep=2pt, leftmargin=1.5em}

\usepackage[english]{babel}
\addto{\captionsenglish}{}

\usepackage{csquotes} 
\usepackage[backend=biber,style=numeric,sorting=none,giveninits=true]{biblatex}
\DeclareSourcemap{
  \maps[datatype=bibtex]{
    \map[overwrite]{
      \step[fieldsource=eprint, final]
      \step[fieldset=url, null]
    }
  }
}
\DeclareSourcemap{
  \maps[datatype=bibtex]{
    \map[overwrite]{
      \step[fieldsource=doi, final]
      \step[fieldset=url, null]
      \step[fieldset=eprint, null]
    }
  }
}
\DeclareSourcemap{
  \maps[datatype=bibtex]{
    \map{
      \step[fieldsource=note, final]
      \step[fieldset=addendum, origfieldval, final]
      \step[fieldset=note, null]
    }
  }
}
\AtEveryBibitem{%
  \clearfield{number}%
  \clearfield{month}%
}
\renewbibmacro{in:}{%
  \ifentrytype{article}{}{\printtext{\bibstring{in}\intitlepunct}}}
\usepackage{etoolbox}
\graphicspath{{figures/}}
\begin{document}

\title{Machine Learning for Screening Large Organic Molecules}
\author{Christopher Gaul \& Santiago Cuesta-Lopez}

\maketitle

\newcommand{\threesubsection}[1]{\emph{#1}.}
\AtBeginEnvironment{tabular}{\fontsize{10}{12}\selectfont}
\AtBeginEnvironment{figure}{\fontsize{10}{12}\selectfont}
\AtBeginEnvironment{tikzpicture}{%
  \fontsize{10}{12}\selectfont%
  \renewcommand{\footnotesize}{\scriptsize}
}

\begin{abstract}
Organic semiconductors are promising materials for cheap, scalable and sustainable electronics, light-emitting diodes and photovoltaics.
For organic photovoltaic cells, it is a challenge to find compounds with suitable properties in the vast chemical compound space. For example, the ionization energy should fit to the optical spectrum of sun light, and the energy levels must allow efficient charge transport.
Here, a machine-learning model is developed for rapidly and accurately estimating the HOMO and LUMO energies of a given molecular structure.
It is build upon the SchNet model (Schütt et al. (2018)) and augmented with a ``Set2Set'' readout module (Vinyals et al. (2016)).
The Set2Set module has more expressive power than sum and average aggregation and is more suitable for the complex quantities under consideration.
Most previous models have been trained and evaluated on rather small molecules.
Therefore, the second contribution is extending the scope of machine-learning methods by adding also larger molecules from other sources and establishing a consistent train/validation/test split.
As a third contribution, we make a multitask ansatz to resolve the problem of  different sources coming at different levels of theory.
All three contributions in conjunction bring the accuracy of the model close to chemical accuracy.
\end{abstract}

\section{Introduction, Motivation, State of the Art}
Organic semiconductors, and in particular organic photovoltaic (OPV) cells, have attracted significant, attention in the last two decades
\cite
{Lussem2013,Firdhaus_2019_NonFullerene,Han_AccChR_2022_NFA-OSC}.
Organic materials for OPV promise to be a cheaper and more scalable approach than silicon.

\subsection{Organic Photovoltaics}
\threesubsection{Brief overview over organic photovoltaics}
The standard setup for organic photovoltaic cells is the bulk heterojunction, i.e., a bicontinuous network of the donor and acceptor materials, as introduced by Yu et al.\ in 1995, with a semiconducting polymer as donor and C\textsubscript{60} (or a derivative) as acceptor \cite{Yu_Science_1995_BHJ}.
After photo excitation, an excited electron can move to the neighboring molecule if there is an unoccupied orbital available at a slightly lower energy.
The small energy difference is needed to overcome the binding energy of the excited electron and the hole it leaves behind.
This charge separation is the principle of OPV.
For high efficiency, it is crucial that the energies of the molecular orbitals match perfectly.

More elaborated OPV cells based on high performance polymer donors and fullerene-based acceptors 
have reached power conversion efficiencies (PCEs) up to 11.7\% \cite{Zhao_NatEnergy_2016}.
More recently, the focus has shifted 
to non-fullerene acceptors (NFAs) \cite{Wadsworth_2019_ReviewNFA,Firdhaus_2019_NonFullerene}, which have reached PCEs over 19\% PCE already \cite{Chong_2022_19percent,Gao_2022_19percent}.

\threesubsection{Materials discovery for OPV}
The power-conversion rates of OPVs are still lower than those of their silicon counterparts, but continuous progress, in particular by discovering better organic materials, has been made in recent years.
In the search for new and better materials, a series of requirements has to be met.
For example, the absorption spectrum should match with the solar spectrum ($E_\mathrm{LUMO} - E_\mathrm{HOMO} \approx \unit[2]{eV}$), 
accompanied with a high extinction coefficient.
The energy levels should be suitably aligned for the intended electrodes and other interface materials (for example at the p-n junction).
The device structure should be stable under various environments, such as humidity, heat, light, and oxygen, and in particular with respect to diffusion (i.e., the molecules should not be too small).
Finally, the materials should be environmental-friendly and non-toxic, and simple fabrication processes, for example, vapor deposition (molecules not too large), should be possible.
The dependence on scarce or expensive raw materials is to be avoided, as far as possible.
Many of these required properties are accessible by ab-initio calculations, i.e., the candidate materials can be simulated before starting experiments in the lab.

\threesubsection{Machine learning in materials discovery for OPV}
Chemical compound space is huge, therefore it is prohibitively expensive to determine the desired properties of each compound with ab-initio methods.
This is where machine-learning (ML) steps in.
Machine learning is a data-driven approach that promises to speed this up, for example by substituting expensive DFT calculations by statistical models that estimate chemical properties at a fraction of the numerical cost.
Figure \ref{fig_big_picture} shows a scheme where a machine-learning algorithm is used as a filter, and only those candidates that pass the ML filter are processed in the subsequent steps (ab-initio calculations and, when applicable, synthetization and characterization in the laboratory).

\subsection{Machine Learning for Quantum Chemistry}

\threesubsection{Bridging the gap between molecular properties and the performance of a OPV device}
A number of efforts has been made to relate the performance of a OPV device to the elemental properties of the materials.
The Scharber model \cite{Scharber_AdMa_2006_BulkHeterojunction} relates the power-conversion efficiency to the donor band gap (under some assumptions on the acceptor and electrode materials).
Padula et al.\ \cite{Padula_2019_PredictOrganicSolarCells,Padula_2019_DonorAcceptorOptimization} tried to go beyond that with a machine learning model taking both electronic and structural features as input.

\threesubsection{Goal and context of present work}
In the present work, we take an intermediate step and develop a ma\-chine-learning model to reliably estimate the HOMO and LUMO energies of a given molecular structure.

\begin{figure}
\definecolor{icamcylred}{HTML}{931d52}
\definecolor{icamcylyellow}{HTML}{fabb00}
\centering
\begin{tikzpicture}[
  scale=0.98,
  draw=black, thick,
  main/.style={draw=icamcylred, very thick},
  grayed/.style={draw=gray},
]
\node (pool) at (0,0) [ellipse, minimum height=2em, main]{
  \parbox{4.2em}{
    \centering pool\\ of molecules}};
\node (ml) at (2.6,0) [isosceles triangle,fill=icamcylyellow!30
, main] {
  \parbox{4.2em}{
    \mbox{ML filter}\footnotesize\\
    \mbox{- \emph{estimate}}\\
    \mbox{\ \ properties}}};
\node (good) at (6.6,0) [rectangle,main] {
  \parbox{5em}{
    \centering ``Good''\\Candidates}};
\draw[->] (pool) -- (ml);
\draw[->] (ml) -- node [above] {\scriptsize accept\quad\ } (good);
\node (trash) at (5.5,-2.2) {\includegraphics{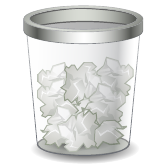}};
\draw[<-,grayed] (trash) +(-0.6,0.9) -- node [left, above, rotate=-28] {\scriptsize \ \ discard} (ml);
{
  \node (ga) at (3.2,2.5) [rectangle, rounded corners, draw]{\parbox{8em}{\centering Genetic Algorithm\\crossover\\\& mutation}};
  \draw[->] (good) .. controls +(0,2) and +(3,0) .. (ga);
  \draw[->] (ga) .. controls +(-3,0) and +(0,2) .. (pool);
}
{
  \node (dft) at (9,0) [isosceles triangle, draw] {
    \parbox{4.2em}{
      \mbox{Ab~Initio}\footnotesize\\
      \mbox{- DFT: \emph{simulate}}\\
      \mbox{\ \ properties}}};
  \node (lab) at (12.7,0) [isosceles triangle, draw] {
    \parbox{3.6em}{
      \mbox{Laboratory}\footnotesize\\
      \mbox{- invest.\ structure}\\
      \mbox{- \emph{measure} props.}\\
      \mbox{- \ \ldots}}};
  \node (prod) at (16.8,0) [rectangle, draw] {\parbox{4.4em}{\centering Enhanced\\Products}};
  \draw[->] (good) -- (dft);
  \draw[->] (dft) -- node [above] {\scriptsize accept\quad\ } (lab);
  \draw[->] (lab) -- node [above] {\scriptsize accept\quad\ } (prod);
}
{
  \node (db) at (8,-3.8) [rectangle, draw,minimum height=3em] {\parbox{5em}{\centering Materials\\Database}};
  \draw[->] (dft)  .. controls +(1.5,-1.75) and +(2,1) ..  (db);
  \draw[->] (lab)  .. controls +(2,-3) and +(4,0) ..  (db);
  \draw[->] (db)  .. controls +(-4,0) and +(-1.7,-2.7) .. node[below,rotate=-22] {
    \parbox{12em}{\centering use for \\
      training, validation, testing\\[-0.9em]}} (ml);
}
\end{tikzpicture}
\caption{A machine learning (ML) algorithm that estimates molecular properties at very low cost, embedded in the big picture of a materials discovery scheme.}\label{fig_big_picture}
\end{figure}
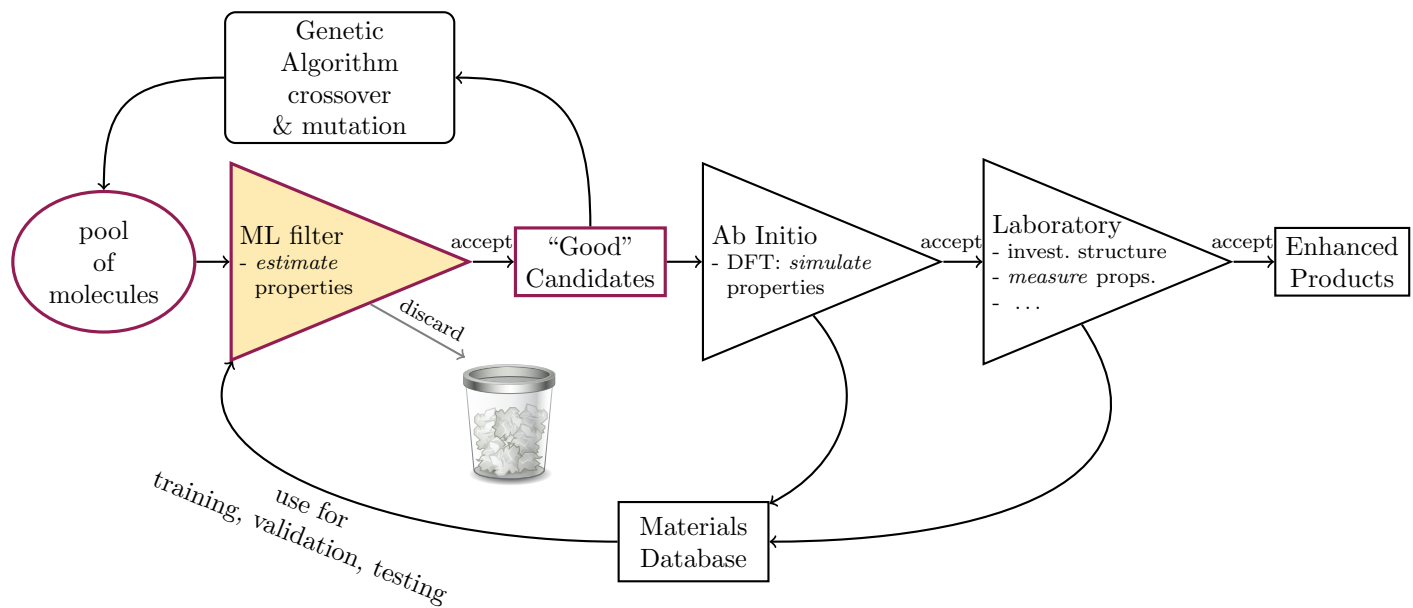

The wider context in a materials-discovery scheme is represented in Fig.\ \ref{fig_big_picture}.
The task of the ML filter is to provide estimates of the molecular properties (currently the HOMO and LUMO energies) in order to identify the ``good'' candidates from a pool of molecule.
Beyond the scope of this article, the good candidates would be further investigated, first by means of ab-initio (usually DFT) calculations, where again, some candidates will be discarded and others pass to the next stage, experiments in the lab, eventually leading to a better product.
Along the way, the knowledge obtained from negative results should not be thrown away, but rather stored in a database in order to be used as training data for future development of the ML algorithm.
Another aspect is applying a genetic algorithm and generate new candidate molecules by means of crossover and mutation of good candidates and feeding them back into the pool of molecules.

\threesubsection{Short review of algorithms for the prediction of molecular properties}
There is a long history of advances in applying ML for predicting chemical properties.
Early approaches used standard ML methods on top if fixed-length molecular fingerprints \cite{Rogers_2010_ECFP}.
But molecular fingerprints are hand-crafted and not invertible.

Duvenaud at al.\ \cite{Duvenaud_NIPS_2015_GraphConvNet}, as well as Kearnes et al.\ \cite{Kearnes_2016_MolecularGraphConvolutions} integrated the fingerprint feature extraction into the deep neural network, which resulted in graph convolutional networks that operate on vertices (atoms) and edges (bonds) of a molecular graph.
This was developed further by Schütt et al.\ \cite{Schuett2018_SchNet}, who added continuous-filter convolutions, and Chen et al.\ who developed MEGNet \cite{Chen_2019_MEGNet}.

Refs.\ \cite{Faber_JTCT_2017_MolecularMachineLearning,Gilmer_2020_ChapterMPNN} are benchmarks of state-of-the art models on QM9 data and have shown that accuracies better than the so-called chemical accuracy ($\unit[1]{kcal/mol} = \unit[0.043]{eV}$) can be reached with models like enn-s2s, SchNet, SchNetE on a dataset of small molecules (for details on QM9, see the following section \ref{sec_data_sources}).

For the real-word application of materials discovery for organic photovoltaics, one has to go beyond the scope of the QM9 dataset.
The contribution of this Article is developing a relevant data corpus (Section \ref{sec_data}) and a suitable architecture for predicting the HOMO and LUMO energies of large organic molecules (Section\ \ref{sec_architecture}), and finally the empiric validation (Section\ \ref{sec_results}).

\section{Data Corpus}\label{sec_data}
\subsection{Data Sources}\label{sec_data_sources}
Several datasets containing significant numbers of molecular structures and properties are already publicly available \cite{Ramakrishnan_2014_QM9,Chen_2019_Alchemy,Stuke_ScientificData_2020_OE62,Lopez_ScientificData_2016_HOPV,Kuzmich_2017_TrendsNFA}.
These sets consists of pairs $(x, y)$, where $x$ is the point cloud of the atoms in space, i.e., a set of quadruples representing the coordinates and the atomic charge of each atom, and $y$ is a set of scalar properties.
The geometries and properties are the outcome of DFT calculations at a given level of theory.
The properties $y$ include, in particular, the energies of the molecular frontier orbitals HOMO and LUMO, which are proxies for the ionization energy and the electron affinity of the respective molecule.

\begin{table}
\centering
\begin{threeparttable}[b]
\caption{Summary of dataset sizes and properties.}\label{tab_data_summary}
\begin{tabular}{lrrccp{11.5em}p{12.6em}}
\toprule
  & & & & & \multicolumn{2}{c}{Level of theory} \\
\cmidrule[0.5pt]{6-7}
Dataset & $N_\mathrm{items}$ & $N_\mathrm{InChIs}$
        & $N_\mathrm{atoms}^\mathrm{median}$ & $N_\mathrm{atoms}^\mathrm{max}$
             & geometry & properties \\
\midrule
QM9 \cite{Ramakrishnan_2014_QM9}
        & 133879 & 129434 & 18 & 29
             & B3LYP/6-31G(2df,p) & \textbf{B3LYP/6-31G(2df,p)} \\
\midrule
Alchemy \cite{Chen_2019_Alchemy}
        & 202579 & 201542 & 22 & 38
             & B3LYP/6-31G(2df,p) & \textbf{B3LYP/6-31G(2df,p)}\\
\midrule
OE62 \cite{Stuke_ScientificData_2020_OE62}
        &  61489 &  61489 & 39 & 174
             & PBE+vdW\tnote{a} / Tier2(tight)\tnote{b}
                     & PBE / Tier2(tight)\\
        &        &        & & &
                     & \textbf{PBE0\tnote{c} / Tier2(tight)}\\
\cmidrule[0.5pt]{2-7}
        &   31k  &    31k & 38 & 116
             & PBE+vdW / Tier2(tight) & PBE0 (water)\tnote{d} / Tier2(tight)\\
\cmidrule[0.5pt]{2-7}
        &    5k  &     5k & 38 & 100
             & PBE+vdW / Tier2(tight) & PBE0 / def2-TZVP\\
        & & & & &    &  PBE0 / def2-QZVP\\
        & & & & &    &  $G_0W_0$@PBE0 / def2-QZVP\\
        & & & & &    &  $G_0W_0$@PBE0 / def2-QZVP\\
\midrule
HOPV \cite{Lopez_ScientificData_2016_HOPV}
        &   4855 &  350  & 69 & 142 & BP86/def2-SVP
                     & BP86/def2-SVP\\
        & & & & &    & \textbf{B3LYP/def2-SVP}\\
        & & & & &    & \textbf{PBE0/def2-SVP}\\
        & & & & &    & M06-2X/def2-SVP \\
\midrule
Kuzmich2017 \cite{Kuzmich_2017_TrendsNFA}
        & 80 & 79\tnote{e} & 123 & 270 & 6-31G*/B3LYP & \textbf{B3LYP/6-31G*}\\
\bottomrule
\end{tabular}
\begin{tablenotes}
\item[a] PBE \cite{Perdew_1996_PBE} level of DFT with TS-vdW corrections \cite{Tkatchenko_2009_TSvdW}.
\item[b] The ``Tier 2 basis'' set in ``tight setting'' is the numeric atom-centered orbitals as implemented in the FHI-aims package \cite{Blum_2009_NAOs}.
\item[c] The hyprid PBE0 exchange correlation functional of Refs.\ \cite{Adamo_1999_PBE0,Perdew_1996_hybrid}.
\item[d] Multipole Expansion (MPE) implicit solvation method for water \cite{Sinstein_2017_Solvation}.
\item[e] Only the molecules with the identifiers FPDI-T and T2 have the same InChI. They differ only by single-bond rotations and have very similar properties.
\end{tablenotes}
Items marked in bold are the ones considered in this work.
\end{threeparttable}
\end{table}

\begin{figure}
\centering
\includegraphics[width=0.3\linewidth]{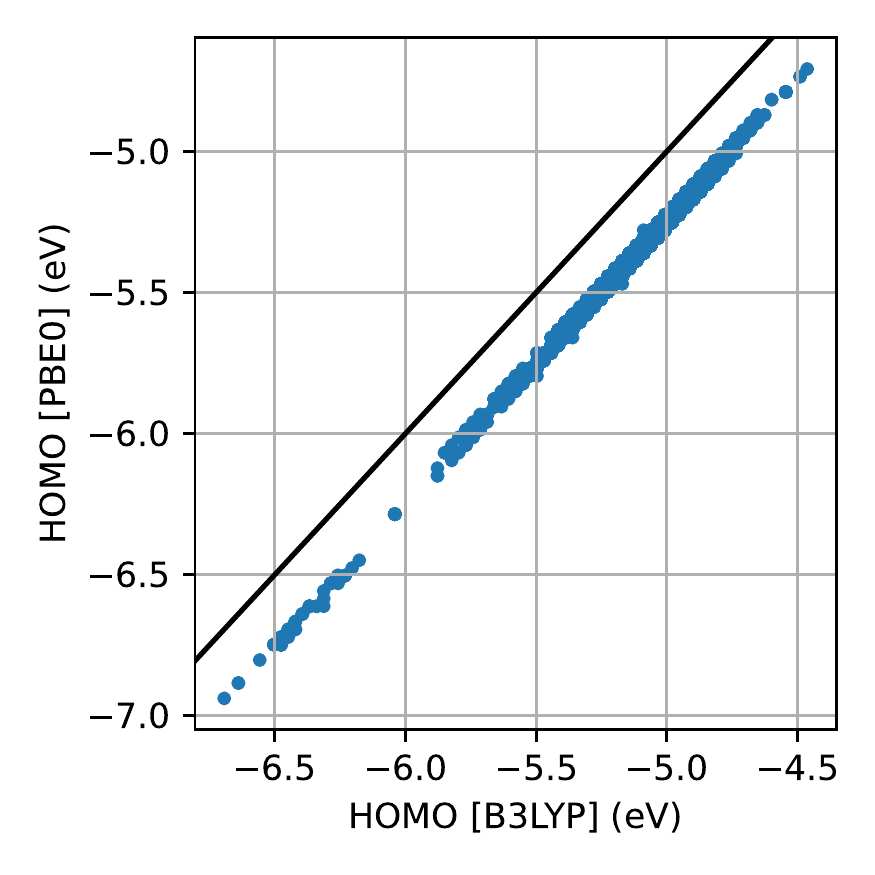}
\caption{Energy of the HOMO obtained by DFT with the PBE0 exchange-correlation functional vs.\ the B3LYP functional for the molecules of the OE62 dataset.
DFT results at different levels of theory are strongly correlated, but may have significant systematic shifts.}\label{fig_PBE0vsB3LYP}
\end{figure}

\begin{itemize}
\item The QM9 dataset \cite{Ramakrishnan_2014_QM9} contains (presumably) all molecules of up to 9 ``heavy'' atoms (C, N, O, F) plus hydrogens, which are 130k structures. The data was obtained at the B3LYP/6-31G(2df,p) level of theory. This dataset is commonly used in many theory papers \cite{Schuett2018_SchNet,Chen_2019_MEGNet,Faber_JTCT_2017_MolecularMachineLearning,Gilmer_2020_ChapterMPNN}.
\item In 2019, Chen et al.\ released a dataset of 200k molecules of up to 12 heavy atoms (C, N, O, F, S and Cl), within the scope of the ``Alchemy'' contest \cite{Chen_2019_Alchemy}.
The level of theory and the set of properties are the same as QM9, but the dataset is extended to slightly larger molecules and larger variety of chemical elements.
\item Stuke et al.\ released the ``OE62'' dataset containing 62k structures, including larger ones \cite{Stuke_ScientificData_2020_OE62}.
The initial structures were extracted from organic crystals in the Cambridge Structural Database (CSD) and the molecular geometry and properties were determined via DFT.
The chemical composition is much wider than those of the other sources, ranging from hydrogen to iodine (H, Li, B, C, N, O, F, Si, P, S, Cl, As, Se, Br, Re, I).
\item The HOPV dataset contains 350 different molecules (5k conformers), which are large and relevant for organic photovoltaics \cite{Lopez_ScientificData_2016_HOPV}. The HOPV set contains the elements
H, C, N, O, F, Si, S, Se.
\item Kuzmich et al.\ published a set of 80 structures that are relevant for organic photovoltaics
\cite{Kuzmich_2017_TrendsNFA}.
Two molecules were discarded because of ambiguous labeling.
Since this set is small, we will use it for testing only.
The variety of chemical elements in the Kuzmich set is the same as HOPV plus B and Cl, which are only in OE62.
\end{itemize}
Table \ref{tab_data_summary} summarizes the properties of the different data sources.
In particular, we will focus on the subset marked in bold face and group them into (i) ``B3LYP'' exchange-correlation functional with a reasonably large basis set (QM9, Alchemy, HOPV) and (ii) ``PBE0'' exchange-correlation functional with a reasonably large basis set (OE62, HOPV).
Ideally, a sufficiently ``good'' level of theory will produce data close to the real values. But in practice, there may be systematic shifts between data obtained from different levels of theory (Figure \ref{fig_PBE0vsB3LYP}).
We will address this problem with a model that makes separate predictions for B3LYP and PBE0, see Sec.\ \ref{sec_multitask} below.

\subsection{Train/Validation/Test Split}

\newcommand\trainrep[8]{%
\begingroup 
\setlength\arraycolsep{1pt}
$\begin{pmatrix}
#1 & #2 & #3 & #4 \\
#5 & #6 & #7 & #8 \\
\end{pmatrix}$
\endgroup}
\newcommand{\ph}{\phantom{O}}

\begin{figure}[tb]
\centering
\includegraphics[width=3cm]{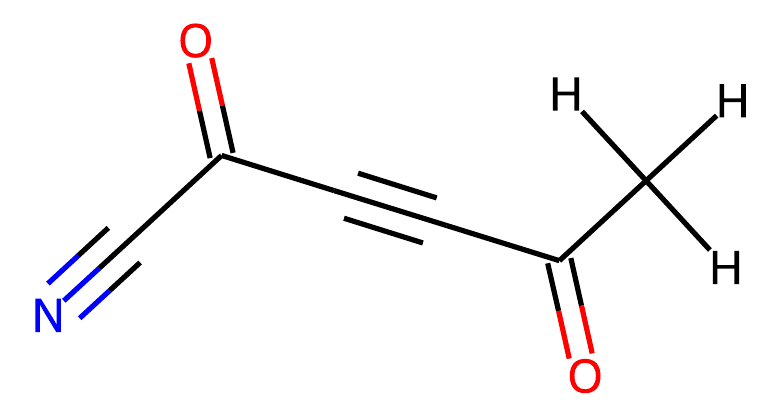}
\caption{The C$_6$H$_3$NO$_2$ molecule (InChI=1S/C6H3NO2/c1-5(8)2-3-6(9)4-7/h1H3), which occurs both in QM9 and Alchemy, is assigned to the training set in both cases.}\label{fig_C6H3NO2}
\end{figure}

In machine learning, the available data is commonly split into three parts, (i) the training data for fitting the free parameters of the model, (ii) the validation data used to make decisions, e.g., when to stop the training process, and (iii) the test set to finally evaluate the accuracy of the model.
For a fair evaluation, it is important that the three subsets are disjoint, in particular, the test shall represent the ability of the model to generalize to data that was not ``seen'' in the training.

Since the same structures may occur in several data sources, care has to be taken that the train/vali\-da\-tion/test split is consistent across all sources.
For example, the molecule C$_6$H$_3$NO$_2$ with the molecular graph as shown in Figure \ref{fig_C6H3NO2} occurs both in QM9 and in Alchemy, like some 18\,000 other molecules.

The international chemical identifier (InChI), which is a unique identifier of a molecule \cite{Heller_2015_InChI} with a layered structure from coarse to fine information.
This opens a convenient way of grouping similar structures together, namely structures that coincide in the first InChI layers.
For example, if only the first two sublayers of standard InChI codes are used, the chemical formula and the backbone structure are the same, but the locations of hydrogens, charge information and the steroechemical information, as well as isotopic information may differ.

The split used in work is constructed according to the following rules:
\begin{enumerate}
\item Perform the split in terms of InChIs truncated after the second layer (molecules that differ only in later layers will be assigned to the same part of the split by construction).
\item Assign the test set of the Alchemy contest to \emph{test}.
\item Assign the original Alchemy validation set to \emph{validation}, except for those structures whose truncated InChIs are already in the test set (2\% of the truncated InChIs), which go to \emph{test}.
\item Assign the rest of the data such that 18\% and 9\% of the truncated InChIs are assigned to \emph{test} and \emph{validation}, respectively. The rest is assigned to \emph{train}.
\item Subsequently assign the truncated InChIs of QM9, OE62, and HOPV in the same manner: Truncated InChIs that have occurred previously are assigned as before; The rest is filled up such that test and validation fraction are 18\% and 9\%, respectively.
\end{enumerate}

\begin{table}
\caption{Intersections of datasets in terms of truncated InChIs (sum formula and non-hydrogen connectivity).
The numbers on the diagonal are the counts of distinct truncated InChIs in each data source.
}\label{tab_trinchis_overlap}
\centering
\begin{tabular}{lrrrr}
\toprule
        &    QM9 & Alchemy &  OE62 &  HOPV \\
\midrule
QM9     & 114707 &   18575 &   340 &     0 \\
Alchemy &  18575 &  174894 &    82 &     0 \\
OE62    &    340 &      82 & 61398 &     3 \\
HOPV    &      0 &       0 &     3 &   347 \\
\bottomrule
\end{tabular}
\end{table}

\begin{table}
\caption{Dataset sizes}\label{tab_dataset_sizes}
\centering
\begin{tabular}{lrrrr}
\toprule
        &  Total &  Train & Valid. & Test \\
\midrule
QM9     & 133879 &  98192 & 12031 & 23656 \\
Alchemy & 202579 & 149644 & 17906 & 35029 \\
OE62    &  61489 &  44895 &  5531 & 11063 \\
HOPV    &   4855 &   3569 &   412 &   874 \\
Kuzmich2017 & 78 &      / &     / &    78 \\
\bottomrule
\end{tabular}
\end{table}

Table \ref{tab_trinchis_overlap} collects the sizes of the different data sets and their intersections in terms of truncated InChIs.
Finally, Table \ref{tab_dataset_sizes} collects the sizes of the train, validation and test parts that result from the procedure described above.

\section{Model Architecture}\label{sec_architecture}
The machine-learning model does not question the training data; it just learns to imitate the results.
As mentioned above (Figure \ref{fig_PBE0vsB3LYP}) that there may be systematic offsets between data obtained at the one or the other level of theory.
Our approach to deal with these systematic shifts between different data sources is that the ML model makes separate predictions for the B3LYP values and the PBE0 values of the frontier orbital energies.

\subsection{Multitask Approach}\label{sec_multitask}
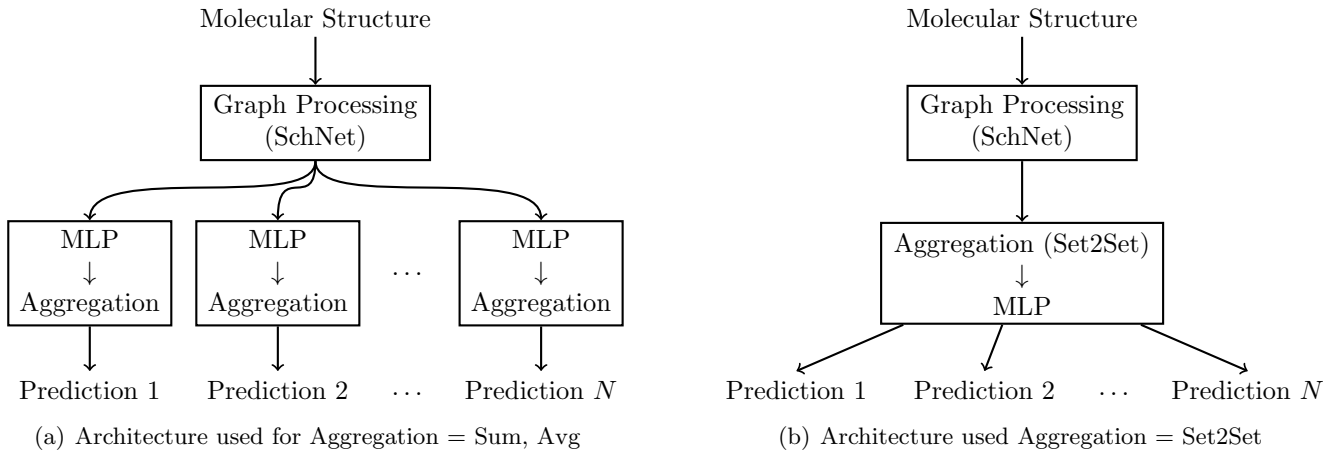
\begin{figure}[bh]
\centering
\subfigure[Architecture used for Aggregation = Sum, Avg\label{fig_architecture_a}]{
\newcommand{\aggrblock}{\parbox{5.5em}{\centering
    MLP\\
    $\downarrow$\\
    Aggregation}}
\begin{tikzpicture}[draw=black, thick]
\node (input) at (0,0) [] {Molecular Structure};
\node (graph) at (0,-1.4) [rectangle,draw] {\parbox{8em}{
  \centering Graph Processing (SchNet)}};
\draw[->] (input) -- (graph);
\node (out1) at (-3,-3.4) [rectangle,draw] {\aggrblock};
\node (graph0) at (0, -2.1) [inner sep=0, outer sep=0] {};
\draw[->] (graph)  .. controls +(0,-1.2) and +(0, 1.5) .. (out1);
\draw[->] (out1) -- + (0, -1.3) node [below] {Prediction 1};
\node (out2) at (-0.5,-3.4) [rectangle,draw] {\aggrblock};
\draw[->] (graph)  .. controls +(0,-1.2) and +(0, 1.5) .. (out2);
\draw[->] (out2) -- + (0, -1.3) node [below] {Prediction 2};
\node at (1.25,-3.4) [] {\ldots};
\node (outN) at (3,-3.4) [rectangle,draw] {\aggrblock};
\draw[->] (graph)  .. controls +(0,-1.2) and +(0, 1.5) .. (outN);
\draw[->] (outN) -- + (0, -1.3) node [below] {Prediction $N$};
\node at (1.25,-5) [] {\ldots};
\end{tikzpicture}
}
\hspace{1.5em}
\subfigure[Architecture used Aggregation = Set2Set\label{fig_architecture_b}]{
\begin{tikzpicture}[draw=black, thick]
\node (input) at (0,0) [] {Molecular Structure};
\node (graph) at (0,-1.4) [rectangle,draw] {\parbox{8em}{
  \centering Graph Processing (SchNet)}};
\draw[->] (input) -- (graph);
\node (out1) at (0,-3.4) [rectangle,draw] {
  \parbox{10em}{\centering
    Aggregation (Set2Set)\\
    $\downarrow$\\
    MLP}};
\draw[->] (graph) -- (out1);
\draw[->] (out1) -- + (-3, -1.3) node [below] {Prediction 1};
\draw[->] (out1) -- + (-0.5, -1.3) node [below] {Prediction 2};
\draw[->] (out1) -- + (3, -1.3) node [below] {Prediction $N$};
\node at (1.25,-5) [] {\ldots};
\end{tikzpicture}
}
\caption{Multitask network architectures.
The multilayer perceptrons (MLP) are small neural networks. }\label{fig_architecture}
\end{figure}

The idea of multitask learning \cite{Caruana1997_multitask} is to train multiple target values in the same model.
The model has a common part and then splits into different branches, one for each \emph{task}. Here, the tasks are the B3LYP estimates and the PBE0 estimates.
Figure \ref{fig_architecture} depicts two example architectures with a single input and multiple outputs (more details given below in sections \ref{sec_graph} and \ref{sec_arch_aggregation}).

A given item from the training set may or may not have target values for all tasks.
For those tasks where targets are available, errors are computed in the loss function and backpropagated through the network.
Notably all tasks contribute to the training of the common part of network and help to learn a meaningful representation of the input.
Since the tasks are correlated, it is expected that they benefit from each other.

\subsection{Molecular-Graph Processing}\label{sec_graph}
\newcommand{\norm}[1]{\lVert#1\rVert}
\newcommand{\rr}{\mathbf{r}}
\newcommand{\x}{\mathbf{x}}
The input to the model are point clouds of atoms, which are readily interpreted as graphs, where the atoms are the vertices and the pairwise distances are the edges.
For the backbone of our model, we use the SchNet architecture by Schütt et al., which was designed to model atomistic systems and makes use of continuous-filter convolutional layers, i.e., learned interactions that respect the symmetries of the underlying physics \cite{Schuett2018_SchNet}.
In a nutshell, SchNet maps the atomic charges to a learnable, $n$-dimensional (here: $n=128$) embedding, which is then processed by several (here: 6) learned interaction modules, each of them incorporating a continuous filter generator.

\threesubsection{The filter generator}
For atoms $i$, $j$, the (continuous) distance $r=\norm{\rr_j - \rr_i}$ is Gaussian expanded,
$e_k = \exp\left(-\gamma(r - \mu_k)^2\right)$, for equidistant $\mu_k$ and $k=0\ldots24$. This expansion passes through two dense layers with shifted-softplus nonlinearities, yielding $W^l_{i, j} \in \mathbb{R}^n$ for the $l$-th filter generator.

\threesubsection{The interaction module}
The atom features are first transformed atom-wisely (matrix multiplication plus adding a bias term).
Then, the continuous-filter operation lets the atoms interact with the other atoms, weighted with the filter,
$\sum_{j=0}^{n_\mathrm{atoms}} \x_j^l \circ W^l_{i, j}$,
where ``$\circ$'' represents the element-wise multiplication.
That is, the interaction module is factorized: the interactions are channel-wise, while channels are mixed in the atom-wise operations.
Then, there are three more layers inside the interaction block, atom-wise, soft-plus nonlinearity, and another atom-wise layer.
Finally, there is a skip connection, i.e., the sum of the outcome of the described interaction chain and its input is returned.

\subsection{Aggregation and Readout}\label{sec_arch_aggregation}
At some point, the representation on a molecular graph of arbitrary size (feature vectors \emph{per atom}) has to be reduced to the set of output estimates, which are \emph{per molecule}.

The simplest way of aggregating would be to take the sum or the average.
This is even the natural choice in some cases.
For example, the energy is an extensive quantity, and the energy of a molecule is the sum of the contributions from all atoms in their local environment.
In fact, this is how the energy is constructed as the sum of local contributions in Ref.\ \cite{Schuett2018_SchNet}.

\threesubsection{Sum and Average Aggregation}
These are the standard aggregation modes in the SchNet paper \cite{Schuett2018_SchNet}.
The (high-dimensional) atomic representation is reduced to the desired output dimension (one for a single regression target) by a small neural network (typically two affine layers with a nonlinearity in-between).
Then, the contributions from the atomic environments are summed up or averaged, which gives the final result, see Figure \ref{fig_architecture_a}.

\threesubsection{Set2Set Aggregation}
The scaling of the present quantities of interest, the HOMO and LUMO energies, with the system size is not clear \emph{a priori}.
Both are related to the level spacing and tend to decrease slowly with the system size, but neither sum nor average is the obvious choice.

Therefore, we resort to the \emph{Set2Set} module that was developed by Vinyals et al.\ \cite{Vinyals_2016_Set2Set} to learn a fixed-length representation for input sets of arbitrary size. It is based on the repeated application of a \emph{long short-term memory} (LSTM) module:
\begin{itemize}
\item The initial LSTM input and hidden (cell) state are zeros.
\item Project the 
  atom features $x$ (one per atom) on the 
  LSTM output $q$ (one per molecule), apply softmax, weighted average $r$ of the atom features $x$.
\item Then, feed $(q, r)$ into the LSTM again.
\item After $N$ repetitions (here, $N=3$), return the vector $(q, r)$, one for each molecule.
\end{itemize}
Thus, the Set2Set module learns to generate a weighted average of the atom features. It has more expressive power than sum or average and, at the same time, respects the invariance with respect to the order of the input features.

The Set2Set aggregation works on the atomic embedding and doubles its dimensions.
Afterwards, a small neural network (here: 2 affine layers with a nonlinearity in-between) reduces the dimensions to the output dimension. Figure \ref{fig_architecture_b} represents the resulting network architecture.

\section{Experiments and Results}\label{sec_results}
\subsection{Impact of Graph Aggregation}\label{sec_aggregation}
To start with, we train a SchNet model on the smallest subset of the training data (QM9 only) with different aggregation modes and output modules (training details and parameters are listed in Sec.\ \ref{sec_training_details}):

\begin{figure}
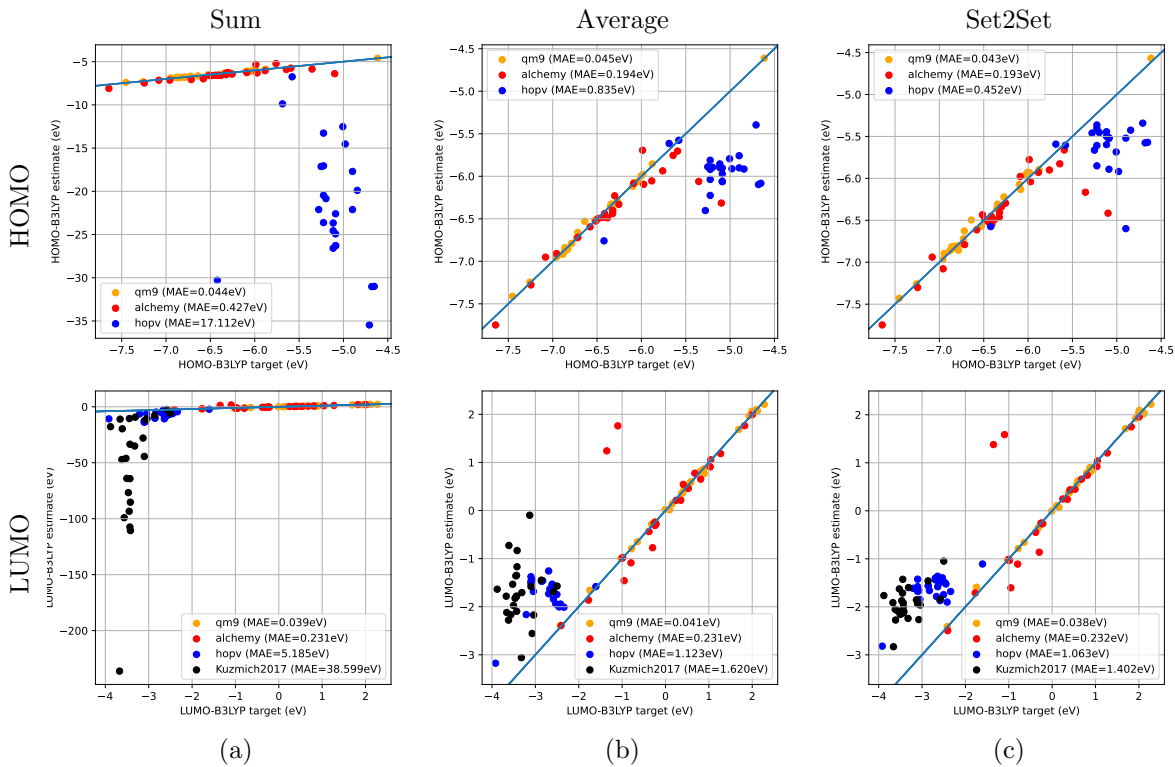

  \newcommand{\accplot}[3]{%
    \includegraphics[width=5cm,clip,trim=0 0 5 42]{#1-#2-#3.pdf}}
  \newcommand{\outputmodulesrow}[1]{
    \raisebox{5em}{\rotatebox{90}{#1}} \qquad
    & \accplot{multitask_model_v08_sum_f}{#1}{B3LYP}
    & \accplot{multitask_model_v08_avg}{#1}{B3LYP}
    & \accplot{multitask_model_v08}{#1}{B3LYP} \\
  }
  \centering
  \setlength{\tabcolsep}{2pt}
  \begin{tabular}{cccc}
    & Sum & Average & Set2Set \\[0.5ex]
    \outputmodulesrow{HOMO}
    \outputmodulesrow{LUMO}
    & (a) & (b) & (c) \\
  \end{tabular}
  \caption{Regression accuracy of SchNet models with different output modules, trained only on the QM9 part of the training data. Shown are the target-estimate pairs of 25 molecules randomly sampled from each test set. The mean average error (MAE) measured on the entire test set is given in the legend.}\label{fig_aggregation}
\end{figure}

\threesubsection{Sum} For each of the targets (HOMO, LUMO, gap), atomic contributions are obtained with a 2-layer neural network and summed up.
Results are shown in column (a) of Fig.\ \ref{fig_aggregation}.
The model achieves good accuracy on test data from the same distribution as the training (QM9), but is completely off for out-of-distribution data, like the large molecules from HOPV and Kuzmich (2017).

\threesubsection{Average} Results for average aggregation (with otherwise the same parameters) are given in panel (b) of Fig.\ \ref{fig_aggregation}.
Average aggregation is not as far off as sum aggregation.

\threesubsection{Set2Set} The set2set output module aggregates in a flexible and learned manner, the dimension is reduced after aggregation.
This yields even better results (panel (c) of Fig.\ \ref{fig_aggregation}) than average aggregation.
Of course systematic deviations of the order of \unit[1]{eV} are still not acceptable, but Set2Set is the best starting point for the next stage, extending the training set.

\subsection{Impact of the Training Data}\label{sec_training_set}
\begin{figure}
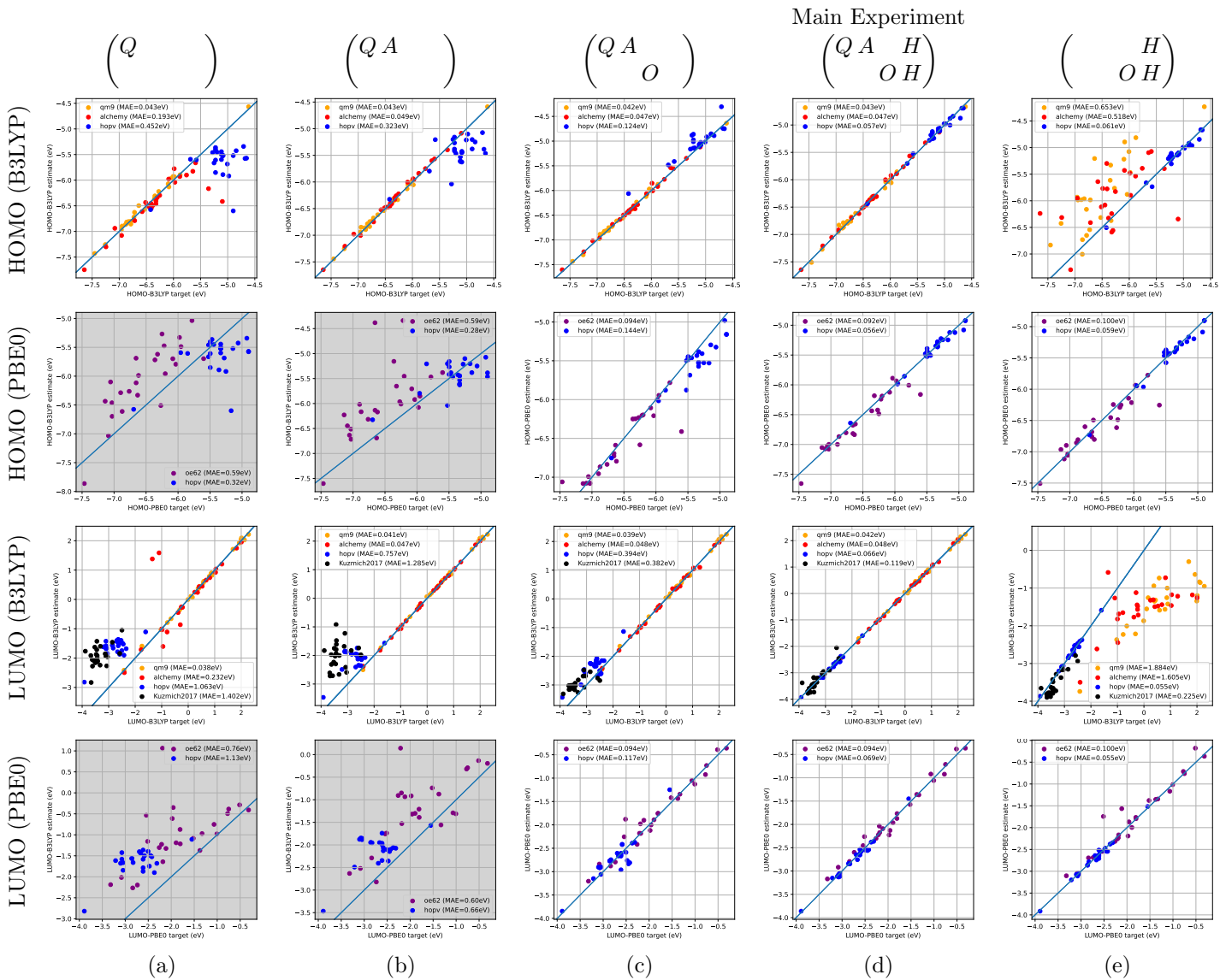

  \newlength{\hght}
  \setlength{\hght}{3.2cm}
  \newcommand{\accplot}[3]{%
    \includegraphics[height=\hght,clip,trim=0 0 5 42]{#1-#2-#3.pdf}%
  }
  \newcommand{\bthreelyprow}[1]{
    \rotatebox{90}{\parbox{\hght}{\centering \quad #1 (B3LYP)}} \qquad
    & \accplot{multitask_model_v08}{#1}{B3LYP}
    & \accplot{multitask_model_v06}{#1}{B3LYP}
    & \accplot{multitask_model_v07}{#1}{B3LYP}
    & \accplot{multitask_model_v01}{#1}{B3LYP}
    & \accplot{multitask_model_v05}{#1}{B3LYP}
    \\
  }
  \newcommand{\pbezerorow}[1]{
    \rotatebox{90}{\parbox{\hght}{\centering \quad #1 (PBE0)}} \qquad
    & \accplot{multitask_model_v08}{#1}{PBE0-cross}
    & \accplot{multitask_model_v06}{#1}{PBE0-cross}
    & \accplot{multitask_model_v07}{#1}{PBE0}
    & \accplot{multitask_model_v01}{#1}{PBE0}
    & \accplot{multitask_model_v05}{#1}{PBE0}
    \\
  }
  \centering
  \setlength{\tabcolsep}{2pt}
  \begin{tabular}{cccccc}
    & & & & Main Experiment & \\[0.2ex]
    & \trainrep{Q}{\ph}{\ph}{\ph}{}{}{}{}
    & \trainrep{Q}{A}{\ph}{\ph}{}{}{}{}
    & \trainrep{Q}{A}{\ph}{\ph}{}{}{O}{}
    & \trainrep{Q}{A}{\ph}{H}{}{}{O}{H}
    & \trainrep{}{}{}{H}{\ph}{\ph}{O}{H} \\[1em]
    \bthreelyprow{HOMO}
    \pbezerorow{HOMO}
    \bthreelyprow{LUMO}
    \pbezerorow{LUMO}
    & (a) & (b) & (c) & (d) & (e)\\
  \end{tabular}
  \caption{Regression accuracy of SchNet models with Set2Set output modules trained on different subsets of the training data, as indicated above (Q = QM9, A = Alchemy, O = OE62, H = HOPV).
  Note that the panels for the PBE0 values in columns (a) and (b) compare B3LYP estimates to PBE0 target values (gray background) because, without PBE0 targets in the training data, the model cannot make separate PBE0 estimates.}\label{fig_training_set}
\end{figure}

We successively extend the training set by adding successively sets of larger molecules (QM9 $\to$ Alchemy $\to$ OE62 $\to$ HOPV), Fig.\ \ref{fig_training_set} (a) -- (d).
This reduces the significantly regression error on the sets of large molecules (HOPV and Kuzmich2017).

From panels (e) to (d), we go in the opposite direction. Whereas (e) is trained without the small molecules, the errors on the QM9 and Alchemy test sets are large. Adding the datasets of small molecules (d), Improves this without harming the accuracy on the large molecules (only HOPV LUMO is a bit worse in (d); HOPV HOMO, OE62, and Kuzmich LUMO are better).

The accuracy of the full model (Set2Set output module for both tasks, full training set, see Fig.\ \ref{fig_training_set}~(d)) has shown errors of the magnitude \unit[0.05]{eV} for HOMO and LUMO for small molecules (QM9, Alchemy) and \unit[0.1]{eV} for large molecules and molecules with large chemical variety (OE62, HOPV, Kuzmich2017).
We stress that this is already close to the so-called ``chemical accuracy'' (1 kcal/mol = 0.043eV) and should be good for practical applications.

\subsection{Ablation Study}

\begin{table}
\caption{Comparing the multi-task model ``Main Experiment'' and the single-task models (B3LYP only and PBE0 only) on the basis of the error measures mean error (ME), mean absolute error (MAE), root-mean-square error (RMSE), all measured in eV.
Bold: best MAE or RMSE value in a given row.
Red: error measures above \unit[0.1]{eV}.}\label{tab_multitask}
\centering
\newcommand{\ME}{ME\phantom{r}}
\newcommand{\MAE}{MAE}
\newcommand{\RMSE}{RMSE}
\newcommand{\g}{\bf}
\renewcommand{\b}{\color{red}} 
\begin{tabular}{llrccrccrcc}
\toprule
 &  & \multicolumn{3}{c}{Main Experiment} & \multicolumn{3}{c}{\phantom{$-$}B3LYP only} & \multicolumn{3}{c}{\phantom{$-$}PBE0 only} \\
Quantity & Test & \ME & \MAE & \RMSE & \ME & \MAE & \RMSE & \ME & \MAE & \RMSE \\
\midrule
\multirow{3}{5em}{HOMO (B3LYP)}
& QM9 & 0.006 & \g0.043 & \g0.063 & 0.008 & 0.046 & 0.068 \\
& Alchemy & 0.001 & \g0.047 & \g0.077 & $-$0.002 & 0.052 & 0.082 \\
& HOPV & 0.004 & \g0.057 & \g0.090 & $-$0.008 & 0.084 & \b0.127 \\
\midrule
\multirow{4}{5em}{LUMO (B3LYP)}
& QM9 & 0.003 & \g0.042 & \g0.062 & 0.002 & \g0.042 & \g0.063 \\
& Alchemy & 0.003 & \g0.048 & \g0.086 & $-$0.001 & 0.050 & 0.090 \\
& HOPV & $-$0.047 & \g0.066 & \g0.088 & $-$0.046 & \b0.115 & \b0.181 \\
& Kuzmich2017 & 0.010 & \g\b0.119 & \g\b0.162 & \b0.479 & \b0.518 & \b0.661 \\
\midrule
\multirow{2}{5em}{HOMO (PBE0)}
& OE62 & 0.001 & \g0.092 & \g\b0.135 & & & & 0.001 & \b0.101 & \b0.148 \\
& HOPV & $-$0.001 & \g0.056 & 0.081 & & & & $-$0.010 & \g0.056 & \g0.072 \\
\midrule
\multirow{2}{5em}{LUMO (PBE0)}
& OE62 & 0.009 & \g0.094 & \g\b0.142 & & & & 0.012 & 0.097 & \b0.146 \\
& HOPV & $-$0.047 & 0.069 & 0.095 & & & & $-$0.014 & \g0.058 & \g0.083 \\
\bottomrule
\end{tabular}
\end{table}

Three efforts have lead to the accuracy achieved in the full model (Fig.\ \ref{fig_training_set}~(d)): the concept of multitask training, the choice of the output module and the extended training set.
Previously, in Sec.\ \ref{sec_training_set}, we have analyzed the impact of the training set, given the design decisions ``multi-task training'' and ``set2set aggregation''.
In this section, we verify that these design decisions were actually important to achieve the accuracy reported above.

\subsubsection{Multi-task learning}
In Table \ref{tab_multitask}, the column ``Main Experiment'' is the multitask model, i.e., B3LYP and PBE0 tasks trained a set2set output module, on top of the representation learned by a common SchNet graph-processing network, as represented in Figure \ref{fig_architecture_b} (with $N = 2 \times 3$ predictions because 3 outputs (HOMO, LUMO and gap) are trained for two tasks, B3LYP and PBE0).
In the other columns, B3LYP task and PBE0 task are trained separately (Figure \ref{fig_architecture_b} with $N=3$ predictions).

For most quantities, the multi-task model achieves better accuracy.
The only exception are the HOPV-PBE0 errors, which are lower in the model trained with PBE0.
But on the other hand, those quantities that are hard in the single-task setting (error measure larger than \unit[0.1]{eV}, marked in red in the columns ``B3LYP only'' and ``PBE0 only''), the errors reduce significantly when the tasks are trained together (column ``Main Experiment'').
We conclude that the concept of multitask-training is indeed important for the accuracy and reliability of the model.

\begin{table}
\caption{Measures of the regression error (in eV) for different output modules.
Bold: best MAE or RMSE value in a given row.
Red: error measures above \unit[0.1]{eV}.
The ``Main Experiment'' is SchNet(6) + Set2Set trained on the full training set.}\label{tab_ablation}
\centering
\newcommand{\ME}{ME\phantom{r}}
\newcommand{\MAE}{MAE}
\newcommand{\RMSE}{RMSE}
\newcommand{\g}{\bf}
\renewcommand{\b}{\color{red}} 
\begin{tabular}{llrccrccrcc}
\toprule
 &  & \multicolumn{3}{c}{Main Experiment} & \multicolumn{3}{c}{\phantom{$-$}SchNet(6) + Average} & \multicolumn{3}{c}{\phantom{$-$}SchNet(6) + Sum} \\
Quantity & Test & \ME & \MAE & \RMSE & \ME & \MAE & \RMSE & \ME & \MAE & \RMSE \\
\midrule
\multirow{3}{5em}{HOMO (B3LYP)}
& QM9 & 0.006 & \g0.043 & \g0.063 & 0.004 & \g0.043 & \g0.064 & 0.008 & 0.049 & 0.079 \\
& Alchemy & 0.001 & \g0.047 & \g0.077 & 0.000 & \g0.048 & \g0.076 & 0.004 & 0.055 & 0.085 \\
& HOPV & 0.004 & \g0.057 & \g0.090 & 0.006 & 0.071 & \b0.108 & $-$0.070 & \b0.222 & \b0.368 \\
\midrule
\multirow{4}{5em}{LUMO (B3LYP)}
& QM9 & 0.003 & \g0.042 & 0.062 & 0.001 & \g0.041 & \g0.059 & 0.010 & 0.046 & 0.068 \\
& Alchemy & 0.003 & \g0.048 & \g0.086 & $-$0.001 & \g0.048 & \g0.087 & 0.010 & 0.052 & 0.088 \\
& HOPV & $-$0.047 & \g0.066 & \g0.088 & $-$0.069 & \b0.145 & \b0.262 & $-$0.087 & \b0.167 & \b0.325 \\
& Kuzmich2017 & 0.010 & \b\g0.119 & \b\g0.162 & \b0.107 & \b0.199 & \b0.240 & \b$-$1.572 & \b1.697 & \b2.363 \\
\midrule
\multirow{2}{5em}{HOMO (PBE0)}
& OE62 & 0.001 & \g0.092 & \b\g0.135 & 0.006 & \b0.109 & \b0.155 & $-$0.011 & \b0.152 & \b0.253 \\
& HOPV & $-$0.001 & 0.056 & 0.081 & 0.006 & \g0.051 & \g0.073 & $-$0.063 & \b0.185 & \b0.274 \\
\midrule
\multirow{2}{5em}{LUMO (PBE0)}
& OE62 & 0.009 & \g0.094 & \b\g0.142 & 0.009 & \b0.109 & \b0.158 & 0.005 & \b0.117 & \b0.169 \\
& HOPV & $-$0.047 & \g0.069 & \g0.095 & $-$0.050 & \b0.143 & \b0.258 & \b$-$0.102 & \b0.154 & \b0.318 \\\bottomrule
\end{tabular}
\end{table}

\subsubsection{Aggregation modules}
We have shown above (Sec.\ \ref{sec_aggregation}) that the Set2Set output module is best if the training represents only a small part of the test distribution (QM9 only).
Here we repeat this investigation with the full training set.
Table~\ref{tab_ablation} collects the results.
The columns labeled ``Main Experiment'' is the model SchNet(6) + Set2Set output module.

\threesubsection{Average aggregation} The model with the \emph{average} output modules is competitive on small molecules (QM9 and Alchemy), which are well represented in the training set. It also performs slightly better on the HOMO(PBE0) task on the HOPV test.
However, it performs much worse on those four tests and quantities that are already difficult for the baseline model \emph{Set2Set} (error measure above \unit[0.1]{eV}, marked in red in the table).
Altogether, the \emph{average} output modules has 12 error measures above \unit[0.1]{eV}.

\threesubsection{Sum aggregation}
When moving on from \emph{average} to \emph{sum}, the picture is similar:
\emph{Sum} performs worse than \emph{average} and \emph{set2set} on those tests and quantities that are already difficult for \emph{average} and \emph{set2set}.
In total, the \emph{sum} output module yields 16 error measures (out of 33) that are above \unit[0.1]{eV}.
The estimates of the LUMO values for the Kuzmich (2017) data are systematically shifted: The mean of the ground-truth LUMO values is \unit[$-3.30$]{eV}, but the mean estimated LUMO is \unit[$-4.86$]{eV}, manifesting unphysical scaling of the estimate with the system size.

In conclusion, out of the three output modules considered, the Set2Set output module remains the most reliable one, especially on challenging data.

\section{Conclusions}
\threesubsection{Achievements}
We have achieved a model to predict the energies of frontier orbitals HOMO and LUMO. It is reliable for large molecules, which are relevant for organic electronics and photovoltaics.

\threesubsection{Enhanced SchNet model}
In practice, it is not enough for a model to be good on a particular subspace of chemical compound space, like, for example, the small molecules of the QM9 dataset.
Rather, the reliability of a model is determined by the difficult test data (here: OE62, HOPV and Kuzmich).
There is no point in reducing the test errors on QM9 below chemical accuracy on QM9-like data if the errors on large molecules are much larger.

We have achieved such a model with reasonable errors on the large molecules
(i) by incorporating the Set2Set output module for improved generalization,
(ii) by extending the training set with data from four different sources, and
(iii) by introducing a multitask ansatz to handle the domain shifts due to different levels of theory in different sources.

\threesubsection{Outlook and Future Work}
The method is not restricted organic photovoltaics and is in general transferable to any problem in materials discovery where material properties are available as training data, think for example of organic light-emitting diodes, organic circuits, materials for batteries or supercapacitors.

Our model can be used in future explorations of the chemical compound space in the search for better materials for organic photovoltaics.
As indicated in Fig.\ \ref{fig_big_picture}, we have interdisciplinary approaches in mind that incorporate machine learning for fast prediction, but also algorithms that generate new candidates, as well as ab-initio methods and experiments for further verification and technical exploitation.

Approaches for molecular geometry generation range from
genetic algorithms \cite{Arapan_2018_magnetic,Henault_2020_Genetic}
over fragment-based approaches \cite{FlamShepherd_2022_FragmentBasedDesign}
to autoencoder-based methods \cite{Kusner_ICML_2017_GrammarVariationalAutoencoder,Simonovsky_ICANN_2018_GraphVAE,FlamShepherd_MLST_2021_MPGVAE}
and generative neural networks \cite{Gebauer_NeurIPS_2019_GSchNet,Gebauer_NatComm_2022_ConditionalGSchNet}.
One challenge for using the current model for such applications is that it is trained with the exact molecular geometries as obtained from DFT geometry optimizations.
If the model is to be applied to structures that come from generating algorithm, the geometries will not be exact and may result in unexpected predictions by the model.
Thus, the sensitivity of the model accuracy with respect to noise in the input coordinates needs be investigated. The problem can probably be alleviated by training the model with random noise added to the input geometries.
Note, however, that this would change the regression task from ``\emph{What are the properties of this molecular geometry?}'' to the question ``\emph{Imagine this geometry relaxes to its near-by ground state, what would be its properties?}''

\medskip
\textbf{Acknowledgements} \par 
This work is supported by the European Union's Horizon 2020 research and innovation program under the Marie Sklodowska-Curie grant agreement No 883256.


\appendix

\section{Experimental Section}
This section contains technical details that should facilitate the reproduction of the results presented in this Article.

\subsection{Training Details}\label{sec_training_details}
The common SchNet network is configured with 6 interaction units, learning a 128-dimensional representation of the atomic environments.
The MLPs before/after the aggregation have 2 layers with the size of the hidden-layer being 64 for the Sum and Average architectures (Fig.\ \ref{fig_architecture_a}) and 32 for the Set2Set architecture (Fig.\ \ref{fig_architecture_b}).

The Adam optimizer is with learning-rate decay from $10^{-4}$ to $5\cdot10^{-7}$ (ReduceLROnPlateauHook with decay factor 0.5 and patience parameter = 25, i.e., the learning rate is reduced if the training error has not reduced in the last 25 epochs).
After each pass through the training data (epoch), the regression error is computed on the validation set.
The model with the lowest validation error is kept (early-stopping criterion based on the validation error).
Typically, the best validation accuracy is achieved close to the first or second learning-rate decay.
Training takes approximately 350 epochs (about 5 days on a nvidia Quadro GP100 GPU).

\subsection{Code and Data}\label{sec_code}
The code to reproduce this work is available at \href{https://github.com/chgaul/maltose}{github.com/chgaul/maltose}, DOI: \href{https://doi.org/10.5281/zenodo.7328587}{10.5281/zenodo.7328587}, including the trained models and scripts to download the primary data from their respective sources.
The maltose code contains
\begin{enumerate}
\item the maltose library, which provides the extensions to the SchNet code described in this article,
\item a collection of scripts to execute and analyze the experiments of this article.
\end{enumerate}

\subsubsection{The Maltose Library}
The library is installed as a python module and provides the set2set output module and multitask functionality.
It builds upon the schnetpack package \cite{Schuett_2019_SchNetPack} (at least v1.0.1), which again makes use of the ``Atomic Simulation Environment (ASE)'' \cite{Larsen_2017_ASE} and pytorch-1.8.0.
The implementation of the Set2Set output module depends on the packages pytorch\_geometric and pytorch\_scatter.

\subsubsection{Running experiments}
\threesubsection{Data preparation}
The scripts for for data download and prepare the datasets (i.e., to convert to \texttt{ase} format) are located at \texttt{scripts/primary\_data/} in the maltose repository, as well as the script that creates the unified train/test/validation split.
These scripts depend on the \texttt{xyz2mol} code from \url{https://github.com/jensengroup/xyz2mol}, which implements the algorithm from Ref.\ \cite{Kim_2015_StructureConversion}, and on the chemical toolkit RDKit \cite{Landrum_2021_rdkit_2021.09.2}

\threesubsection{Configuration and execution of experiments}
The experiment configurations are provided in \texttt{scripts/configs/}.
There are scripts to run the training, to evaluate the errors, and to create the plots included in this article.


\medskip

%
\printbibliography


\end{document}